\documentclass[10pt,twocolumn]{hdthep}
\input epsf

\usepackage{amsmath,amssymb,bm,graphicx,booktabs}
\usepackage{epsfig}  
\usepackage{subfigure}

\newcommand{\lsim}{\lower.7ex\hbox{$\;\stackrel{\textstyle<}{\sim}\;$}}

\newcommand{\weff}{\ensuremath{\overline{w}_0}}
\newcommand{\ome}[2][]{ \ensuremath{
    \Omega_{\textrm{#1}}^{#2}}} 
\newcommand{\omebar}[2][]{\ensuremath{
    \overline{\Omega}_{\textrm{#1}}^{\ensuremath{#2}}}}
\newcommand{\sh}[1][]{\ensuremath{\delta \varphi_{#1}}}
\newcommand{\osh}[1][]{\ensuremath{\varphi_{#1}}}

\newcommand{\rstar}{\ensuremath{r_\star}}

\newcommand{\obh}{\ensuremath{\Omega_b h^2}}

\newcommand{\pam}{\ensuremath{\left (\obh, r^*, n,
      \overline{\Omega}_{\rm ls}^{\phi} \right)}}
\begin{document}
\hdthep{01-22}
\pubdate{April 2001}
\title{The Location of CMB Peaks in a Universe with Dark Energy}
\author{Michael Doran  and Matthew Lilley}

\abst{
The locations of the peaks of the
CMB spectrum are sensitive indicators of cosmological parameters, yet
there is no known analytic formula which accurately describes their
dependence on them.  We parametrize the location of the peaks as $l_m
= l_A(m - \varphi_m)$, where $l_A$ is the analytically calculable
acoustic scale and $m$ labels the peak number.  Fitting formulae for
the phase shifts $\varphi_m$ for the first three peaks and the first
trough are given.  It is shown that in a wide range of parameter
space, the acoustic scale $l_A$ can be retrieved from actual CMB
measurements of the first three peaks within one percent accuracy.
This can be used to speed up likelihood analysis.  We describe how the
peak shifts can be used to distinguish between different models of
dark energy.}

\maketitle
%\begin{keywords}
%cosmic microwave background---cosmology: theory
%\end{keywords}

\section{Introduction}
The locations of the peaks and troughs of the CMB anisotropy spectrum
can serve as a sensitive probe of cosmological parameters
\cite{Huey:1999se, Hu:1996qz,Amendola:2000er,Brax:2000yb,Coble:1997te,Doran:2000jt}.
%(Huey et. al. 1999; Hu \& White 1996; Amendola 2000; Brax, Martin \& Riazuelo 2000; 
%Coble, Dodelson \& Frieman 1997; Doran et. al. 2000)

There are however many processes which contribute to the final
anisotropies, and these must be calculated from systems of coupled
partial differential equations \cite{Seljak:1996is}.  As such it is
not possible {\em a priori} to derive an accurate analytic formula for
the peak locations.  There exists a numerically-obtained estimate of
the location of the first peak \cite{Kamionkowski:1994aw} for a
universe with no cosmological constant, namely $l_1 \sim 200\ 
\Omega_m^{-1/2}$.  This was recently extended to universes with
$\Lambda \neq 0$, by perturbing around the $\Lambda = 0$ value
\cite{Weinberg:2000ts}, but holding all other parameters fixed.  In
this work, we calculate the locations of the first three peaks as a
function of several cosmological parameters, including universes with
a large dark energy component.  We show how these results can be used
to extract cosmological information about, for instance the history of
quintessence, from just a handful of CMB data points and also to speed
up multi-parameter likelihood analysis.

Before last scattering, the photons and baryons are tightly bound by
Compton scattering and behave as a fluid.  The oscillations of this
fluid, occuring as a result of the balance between the gravitational
interactions and the photon pressure, lead to the familiar spectrum of
peaks and troughs in the averaged temperature anisotropy spectrum
which we measure today.  The odd peaks correspond to maximum
compression of the fluid, the even ones to rarefaction
\cite{Hu:1997qs}.  In an idealised model of the fluid, there is an
analytic relation for the location of the $m$-th peak: $l_m \approx
m\, l_A$ \cite{Hu:2000ti,Hu:1995uz} where $l_A$ is the {\em acoustic
  scale} which may be calculated analytically \cite{Doran:2000jt} and
depends on both pre- and post-recombination physics as well as the
geometry of the universe.

\begin{table}
\begin{center}
\begin{tabular}{ccccc}
\toprule
$\Omega_m$ & $\Omega_\Lambda$ & $l_1$ (estim.) & $l_1$ (numeric.) &
$\%$ error\\
\midrule
0.4 & 0.6 & 296 &219 & 35\\
1.0 & 0.0 & 269 &205 & 31\\
\bottomrule
\end{tabular}
\caption{Values of the location of the first peak $l_1$ estimated by
  $l_1 \approx l_A$ and calculated numerically via {\protect\textsc
    CMBFAST} {\protect\cite{Seljak:1996is}}.  The intuitive model clearly does
  not describe the location of the first peak well, though the
  spacings between other peaks is better.  The above values were
  calculated assuming $h=0.65$, $\Omega_b = 0.05$, $n=1$ and $a_{\rm
    ls} = 1100^{-1}$.}
\label{table:error}
\end{center}
\end{table}

The simple relation $l_m \approx m\, l_A$ however does not hold very
well for the first peak (see Table \ref{table:error}) although it is
better for higher peaks \cite{Hu:1996qz}. Driving effects from the
decay of the gravitational potential as well as contributions from the
Doppler shift of the oscillating fluid introduce a shift in the
spectrum.  In order to compensate for this, we parametrize the
location of the peaks and troughs as in \cite{Hu:2000ti} by\footnote{
  The peaks are labelled by integer values of $m$ and the troughs by
  half-integer values.}
\begin{equation} \label{our_phi}
l_m \equiv l_A \left(m - \osh[m]\right) \equiv  l_A \left(m -
  \bar{\varphi} - \sh[m] \right). 
\end{equation}
For convenience, we define $\bar\varphi \equiv \osh[1]$ to be the
overall peak shift, and $\sh[m] \equiv \osh[m] - \bar\varphi$ the
relative shift of the $m$-th peak relative to the first.  The reason
for this parametrization is that the phase shifts of the peaks are
determined predominantly by pre-recombination physics, and are
independent of the geometry of the Universe.  In particular, the ratio
of the locations of the first and $m$-th peaks
\begin{equation}\label{ratio}
\frac{l_m}{l_1} = \frac{l_A}{l_A} \frac{ \left(m -  
      \bar{\varphi} - \delta \varphi_m \right)}{\left(1 -
      \bar{\varphi}\right)} = 1 + \frac{m-1 -\delta
      \varphi_m }{1 - \bar{\varphi}},
\end{equation}
probes mostly pre-recombination physics and so can be used to extract
information on the amount of dark energy present before last
scattering \cite{Doran:2000jt}.

If we knew how the phase shifts depended on cosmological parameters,
it would be possible to extract $l_A$ from the measured CMB spectrum.
Since any given cosmological model predicts a certain value of $l_A$,
this is a simple way of distinguishing between different models -- in
particular it has been shown \cite{Doran:2000jt} that different
quintessence models with the same energy density and equation of state
today can have significantly different values of $l_A$.  Finally,
having extracted $l_A$ from observations, we could speed up likelihood
analysis by being able to discard models not leading to the right
value of the acoustic scale before a single perturbation equation has
to be solved.

In a recent paper \cite{Hu:2000ti}, a fitting formula
for $\bar{\varphi}$ was given
\begin{equation} \label{phase_fit}
\bar{\varphi} \approx 0.267 \left(\frac{\rstar }{ 0.3}\right)^{0.1},
\end{equation}
for the values $n=1$, $\Omega_b h^2 = 0.02$.  In this formula,
$\rstar$ is the ratio of radiation to matter at last
scattering\footnote{This relation also holds in the presence of
  quintessence.}
\begin{equation}\label{r_star}
\rstar = \rho_r(z_\star) / \rho_m(z_\star) = 0.042 \left(\Omega_m
  h^2\right)^{-1} \left(z_\star / 10^3\right). 
\end{equation}
Equation (\ref{phase_fit}) however, is valid only for the given values
of spectral index, Hubble parameter and baryon density.  It does not
include the dependence of the peak location on the amount of
quintessence present at last scattering, and is valid only for the
first peak $l_1$. In this paper, we give fitting formulae (see Appendix \ref{sec::fit}) 
for the shifts of the first three peaks and the first trough and describe how
one can use them to extract cosmological information from future CMB
experiments.

\begin{table}
\begin{center}
\begin{tabular}{cc}
\toprule
Symbol & Range\\
\midrule
\ome[0]{m} & $[0.2,\,0.6]$ \\ 
\obh  & $[0.005,\,0.04]$ \\  
\omebar[ls]{\phi} & $[0,\,0.23]$ \\ 
$h$ &  $[0.55,\,0.80]$ \\
$n$ &  $[0.8,\,1.2]$ \\ 
\bottomrule
\end{tabular}
\caption{Parameter ranges used in this work.}
\label{tab::parameters}
\end{center}
\normalsize
\end{table}

Our first task in computing fitting formulae for the peak locations is
to decide which cosmological parameters to fit to.  The dependence on
the baryon density and the Hubble parameter is sensitive only to the
product $\Omega_b h^2$, and so we do not seek to fit for them
separately.  We further take $\rstar$ defined in Equation (\ref{r_star}) 
and the spectral index $n$ as parameters.
For the quintessence dependence, we use the effective average
density component before last scattering $\omebar[ls]{\phi}$ defined
as in \cite{Doran:2000jt}
\begin{equation}\label{omebar_defn}
{\omebar[ls]{\phi}} \equiv \tau_{\rm ls}^{-1}
\int_0^{\tau_{\rm ls}} \ome[]{\phi}(\tau) \textrm{d}\tau.
\end{equation}

\begin{figure*}
\begin{minipage}{6.2in}
\begin{tabular}{rr}
\subfigure[]{\includegraphics[scale=0.6]{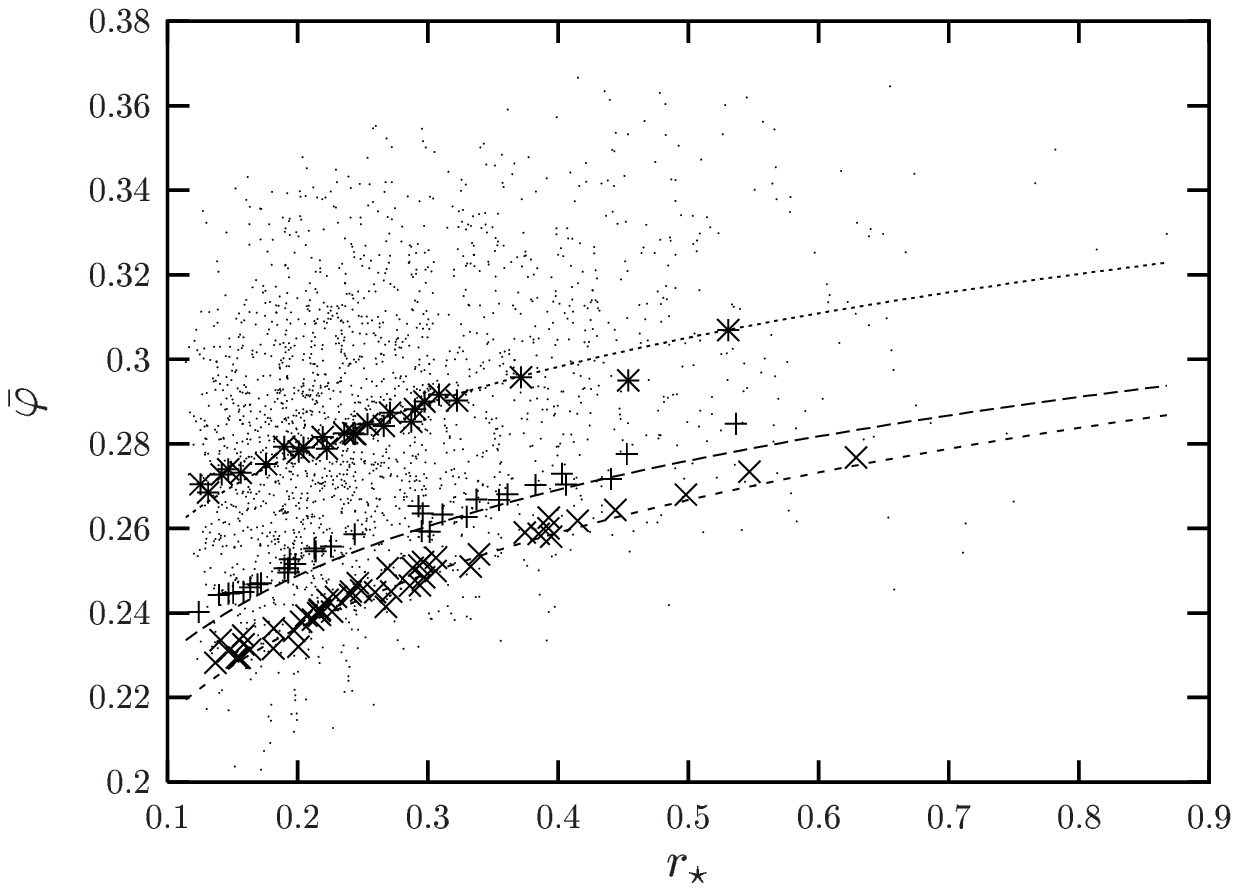}} &
\subfigure[]{\includegraphics[scale=0.6]{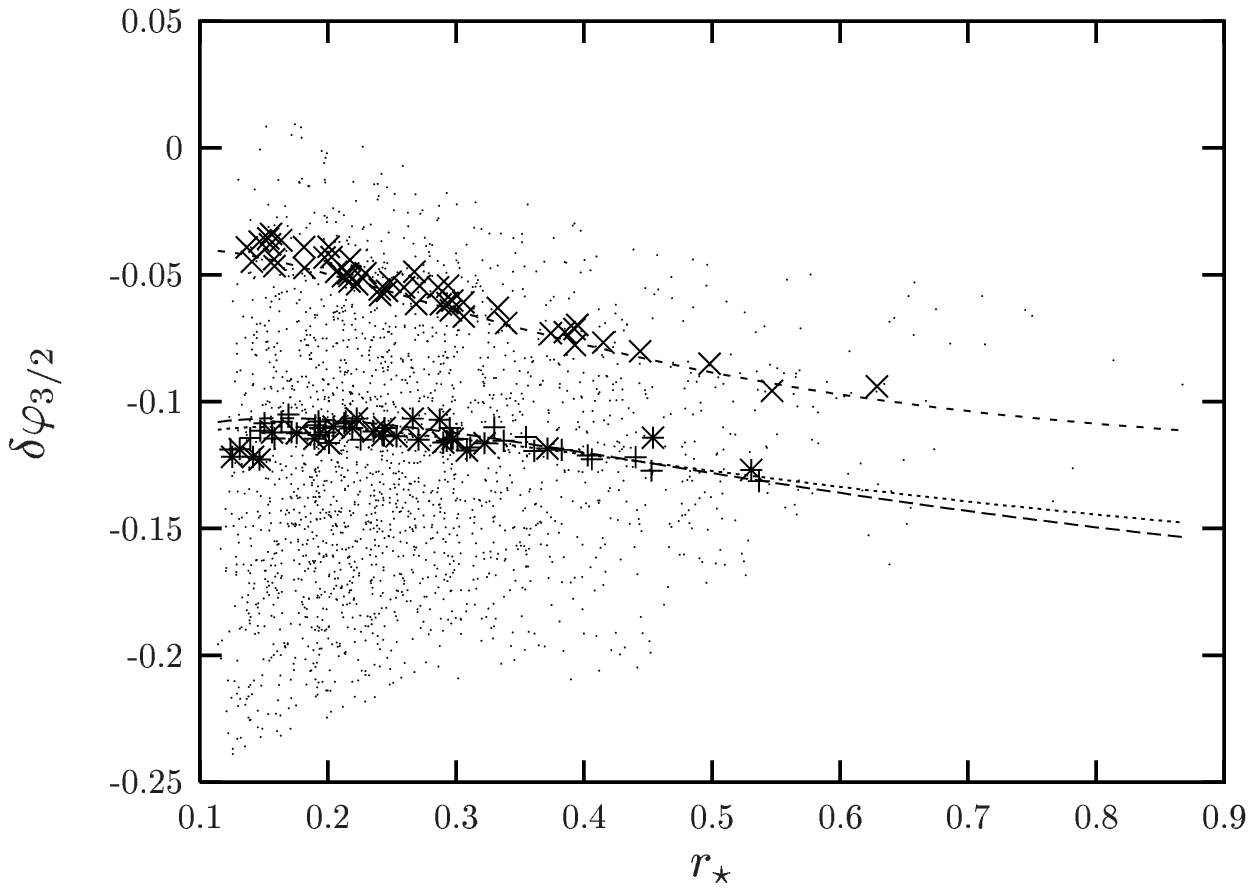}}\\
\subfigure[]{\includegraphics[scale=0.6]{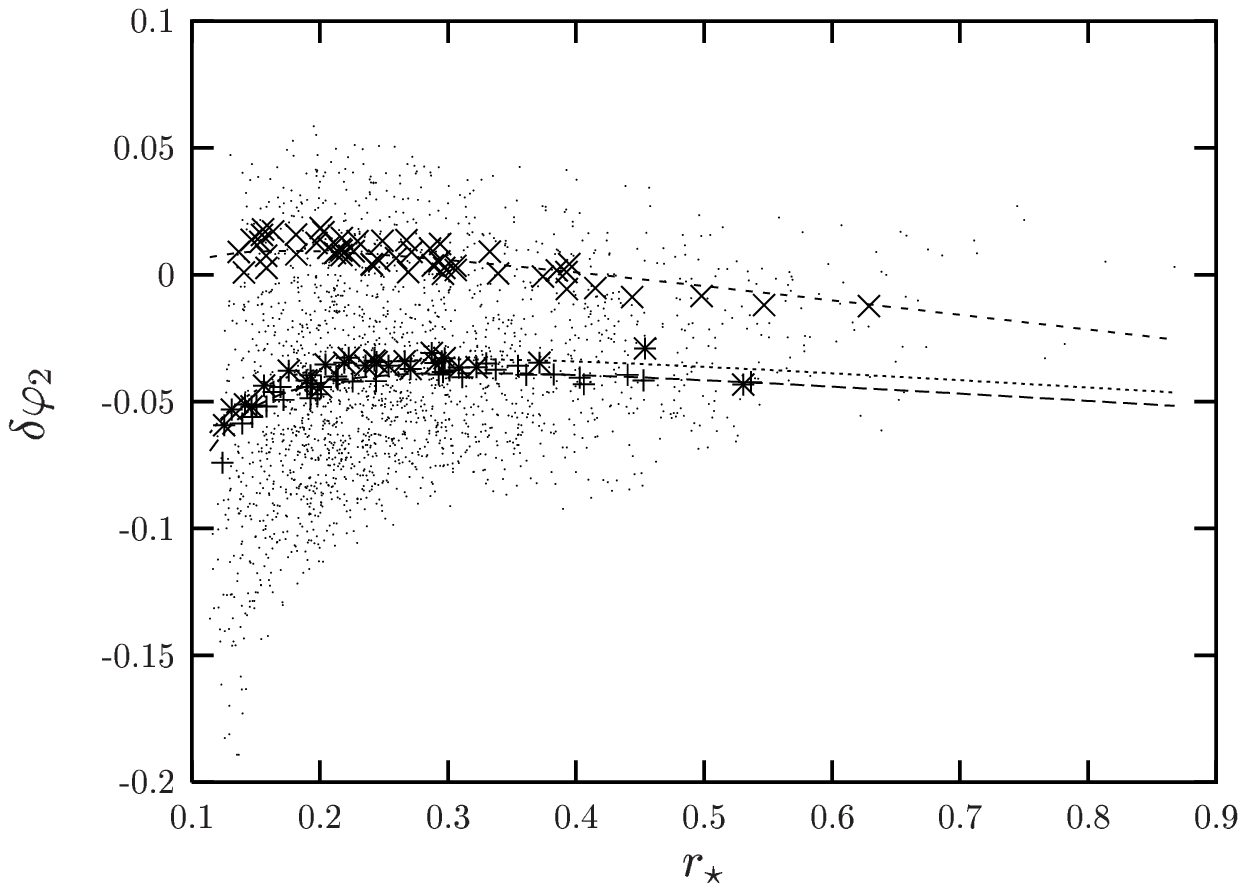}} &
\subfigure[\label{fig::p3o}]{\includegraphics[scale=0.6]{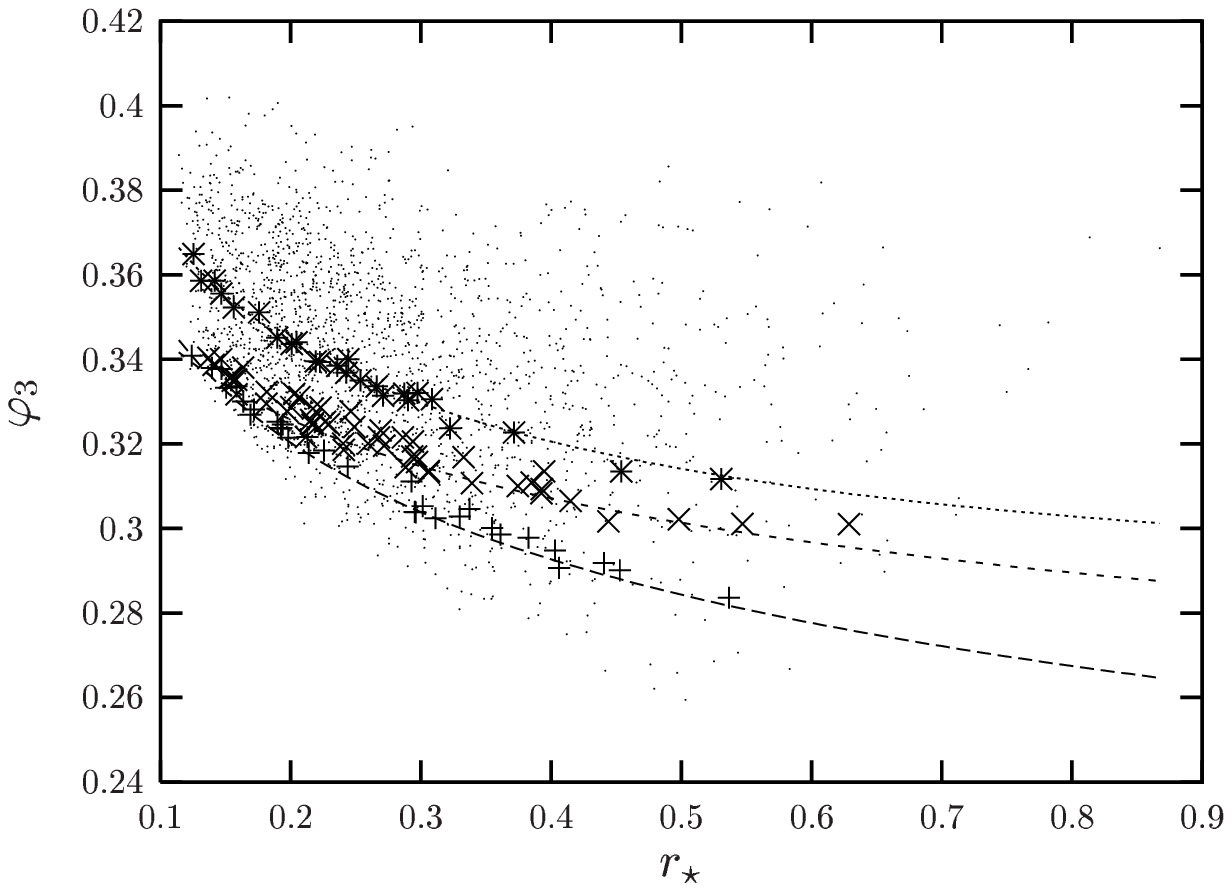}}
\end{tabular}
\label{fig::histphi3}
\caption{The overall shift $\bar\varphi$ (a) and the relative shifts of 
  the first trough (b) and the second peak (c). Also given is the
  overall shift of the third peak (d).  In all figures, the long
  dashed, dotted and the dashed lines represent the fitting formulae
  for the parameters $\pam = (0.02,\,\rstar,\,1,\,0)$,\ 
  $(0.02,\,\rstar,\,1,\, 0.1)$ and $(0.01,\,\rstar,\,1,\,0)$
  respectively.  The large symbols show the data corresponding to these
  curves. The errors quoted in Appendix \ref{sec::fit} are calculated
  from the spread of these symbols relative to the curves.
  The sprinkled dots represent thousands of models
  selected at random from the parameter space given in Table
  \ref{tab::parameters}, and indicate the ranges of  values taken on
  by $\bar\varphi$ etc. for these models.}
\end{minipage}
\end{figure*}

We recall that the peak shifts are sensitive mainly to
pre-recombination physics and so we do not need to use the value of
$\Omega^{\phi}$ today as a parameter.  Of course the acoustic scale
$l_A$ does depend on today's quintessence component -- we give a
relation for $l_A$ in Section \ref{sec::estimate}. We will thus seek
to find the dependence of $(\bar{\varphi},\delta\varphi_m)$ on the
cosmological parameter set \pam.  In performing these calculations, we
restricted each of the cosmological parameters used to lie within a
certain interval, which in each case is over- rather than
under-cautious.  The ranges of parameter values chosen are displayed
in Table \ref{tab::parameters}. To gain intuition for the fitting formulae, 
we plot curves for the shift of the first and the second peak as well
as the relative shifts of the first trough and the second peak in 
Figure \ref{fig::histphi3}.

In Sections \ref{sec::phase} and \ref{sec::estimate} we describe a
systematic procedure for extracting the acoustic scale $l_A$ from the
location of the first three peaks.  Section \ref{sec::kappa}
introduces a quantity $\kappa$ which is useful as it depends only on
two of our four parameters.  
The model (in)dependence of the fitting formulae is discussed in 
Section \ref{sec::dependence}.
Finally, our fitting formulae are given in Appendix
\ref{sec::fit}.

\section{Retrieving the shifts from CMB measurements}
\label{sec::phase}

With future high precision measurements of the {\it MAP}\footnote{http://map.gsfc.nasa.gov/}
and
{\it PLANCK}\footnote{http://astro.estec.esa.nl/SA-general/Projects/Planck/}
satellites, we expect that the position of the
first three peaks and troughs will be determined to high accuracy.
From these few data points, it is possible to extract valuable
information on the cosmological parameters.  We have observed, during
our computation of CMB spectra for thousands of universes, that the
overall shift of the third peak \osh[3] (i.e.  \osh[3] $=
\bar{\varphi}+\sh[3]$) is a relatively insensitive quantity.  In the
parameter range we used (see Table \ref{tab::parameters}) we found
that $\osh[3] =0.341 \pm 0.024$.\footnote{Here and in the following,
  we quote 1-$\sigma$ errors.  All errors follow approximately a bell
  curve.} In using $\osh[3] = 0.341$ we introduce slight (at most one
percent) systematic deviations in our estimate, because an increase of
$\omebar[ls]{\phi}$ typically increases $\osh[3]$ (see Fig.
\ref{fig::p3o}).  We will partially correct for these effects by
improving our estimate for $\osh[3]$, via the procedure described
below.

We start by extracting our first estimate of the overall phase shift,
from the measured locations of the first and third peaks
\begin{equation}\label{est_phi}
\bar \varphi = 1 - (3 - \varphi_3) \frac{l_1}{l_3} \approx  1 - 2.66
\frac{l_1}{l_3}.
\end{equation}
Comparing this estimate with the value calculated from numerical
simulations, we find $\Delta \bar\varphi = 0.006$.  Having a handle on
the overall phase shift, it is now simple to infer the relative shifts
\sh[m] of the remaining troughs and peaks. From equation (\ref{ratio})
we get the relation
\begin{equation}\label{est_dphi}
\sh[m] = (m-1) - \left (\frac{l_m}{l_1} - 1 \right )\left (1 -
  \bar\varphi \right).
\end{equation}
The error of this estimate is 
\begin{equation}
\Delta \left(\sh[m]\right) =   \left (\frac{l_m}{l_1} - 1 \right )
\Delta  \bar\varphi. 
\end{equation}
Having a first (and already quite accurate) estimate of the shifts, we
now correct for the systematic effects described above. Taking the
cosmological parameter set we wish to maximise over (i.e. Table
\ref{tab::parameters}), we calculate for each model universe the phase
shifts of the first three peaks using the fitting formulae given in
Appendix \ref{sec::fit}.  We then discard those models for which any
phase shift deviates significantly (say $>2$-$\sigma$) from the
data-inferred values.  This leaves an improved cosmological parameter
set, for which the average value of $\osh[3]$ is calculated (see
Table \ref{tab::sys3}).  This
improved $\osh[3]$ can then be used to re-calculate the phase shifts
from Equations (\ref{est_phi}) and (\ref{est_dphi}).

\begin{table}
\begin{center}
\begin{tabular}{r@{\,-\,}lcc}
\toprule
\multicolumn{2}{c}
{\omebar[ls]{\phi}  ($\%$) }& $\langle \osh[3]^{\rm num} 
\rangle$ & $\langle \osh[3]^{\rm improved} \rangle$ \\
\midrule
0 & 2 & 0.313 & 0.326 \\
10&12 & 0.340 & 0.337 \\
18 & 20 & 0.362  & 0.348 \\
\bottomrule
\end{tabular}
\caption{Binned average $\osh[3]$ of the numerical simulation and the
  improved deduction.}
\label{tab::sys3}
\end{center}
\normalsize
\end{table}

\section{Estimating $\lowercase{l}_A$}\label{sec::estimate}
Using the improved value\footnote{In fact, using $\osh[3] = 0.34$
  instead of the improved value also gives reasonable results.}  for
$\osh[3]$ from the previous section, we can extract to very good
accuracy the acoustic scale $l_A$ -- the quantity which determines the
overall spacing of CMB
peaks:
\begin{equation}
l_A = \frac{l_3}{3-\varphi_3} % \approx \frac{l_3}{2.66}
\end{equation}
In fact, the deviation of the value of $l_A$ estimated from this
formula and the numerically-obtained value is small, with a
$1$-$\sigma$ error of $0.8\%$ (see also Table \ref{tab::leaping}).
This is a very valuable result, for the value of $l_A$ can be simply
computed for any given quintessence (or indeed any other) cosmology.
For flat universes it is given by \cite{Doran:2000jt}
\begin{multline}
\label{sep}
 l_A =   \pi \bar c_s^{-1} \Bigg[
      \frac{F(\ome[0]{\phi},\weff)}{\sqrt{1-{\omebar[ls]{\phi}}}}
      \Bigg \{ \sqrt{a_{\rm ls} +
      \frac{\ome[0]{r}}{ 1 - \ome[0]{\phi}}} \\
 - \sqrt{\frac{\ome[0]{r}}{1 - \ome[0]{\phi}}} \Bigg \} ^{-1} - 1 \Bigg],
\end{multline}
with
\begin{multline} \label{F_int}
F(\ome[0]{\phi},\weff) = \frac{1}{2} \int_0^1 \textrm{d}a \Bigg( a +
  \frac{\ome[0]{\phi}}{1-\ome[0]{\phi}} \, a ^{(1 - 3 \weff)} \\
  + \frac{\ome[0]{r}(1-a)}{1-\ome[0]{\phi}} \Bigg)^{-1/2},
\end{multline}  
where $\ome[0]{r}, \ome[0]{\phi}$ are today's radiation and
quintessence components, $a_{\rm ls}$ is the scale factor at last
scattering (if $a_0=1$) and  $\bar c_s$ is the average sound speed before
last scattering:
\begin{equation}
\bar c_s \equiv \tau_{\rm ls}^{-1} \int_0^{\tau_{\rm
    ls}} {\rm d}\tau \left [ 3 + (9/4)(\rho^{\rm   b}(t)/\rho^{\gamma}(t))  \right ]^{-1/2}.
\end{equation} 
The effective equation of state, $\weff$ is the $\Omega^\phi$-weighted
average over conformal time
\begin{equation}
\label{w_eff}
\weff =  \int_0^{\tau_0} \ome[]{\phi}(\tau) w(\tau) \textrm{d} \tau
\times \left(  \int_0^{\tau_0} \ome[]{\phi}(\tau) \textrm{d} \tau
\right)^{-1}.
\end{equation} 
In particular, different quintessence models with the same energy
density and equation of state today can have significantly different
values of $l_A$.  In this way stringent bounds on cosmological models
can be imposed just by comparing the $l_A$ value of specific models.

\section{Insensitive Quantities} \label{sec::kappa}

\begin{figure}
\scalebox{0.6}[0.6]{
\includegraphics{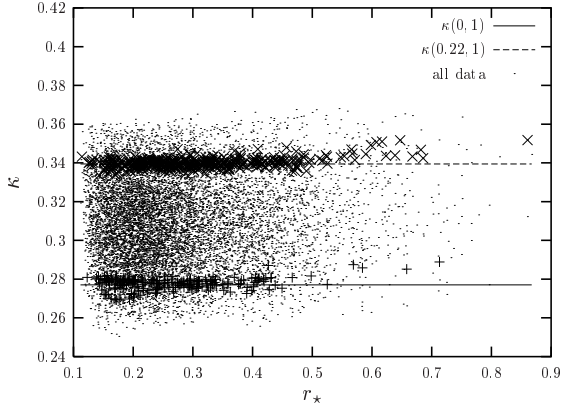}}
\caption{The quantity $\kappa$
    as a function of $\rstar$. It is practically insensitive to
    $\rstar$ and $\obh$ for most of the initial conditions considered.
    The dots represent fifty thousand models with parameters in the
    ranges given in Table \ref{tab::parameters} The $+$'s and
    $\times$'s represent models with \omebar[ls]{\phi} = 0 and 0.22
    respectively, for $n=1$, and all values of other input parameters.
    \label{fig::kappaOverview}}
\end{figure}

The phase shifts depend on the cosmological parameters \pam. Of
course, if it were possible to find a linear combination of phase
shifts which is insensitive to some of these parameters and thus
reduce the dimensionality of our parameter space, it would greatly
help in extracting cosmological information.  To this end, we note an
anti-correlation between $\bar\varphi$ and \sh[3] -- empirically, we
have found that the quantity
\begin{equation}
\kappa \equiv \bar\varphi + \frac{2}{5} \sh[3]
\end{equation}
is practically insensitive to $\rstar$ and $\obh$, and depends only on
$n$ and $\omebar[ls]{\phi}$. In fact, it is to very good approximation
given by the fit
\begin{equation}
\kappa = \left (0.277 + 0.284  \omebar[ls]{\phi} \right ) (1.3 -
0.3 n),
\end{equation}
with $\Delta \kappa^{\rm fit} \approx 0.0024$ being the deviation of
the fit from the numerically-simulated values (see Fig.
\ref{fig::kappaOverview}).  Following the procedure in Section
\ref{sec::phase}, we can deduce $\kappa$ from the measured values of
the peak locations. Within our parameter range, $\kappa$ is then
determined with error $\Delta \kappa ^{\rm deduc.} = 0.013$.

In the parameter space we have considered, the value of $\kappa$
varies between $0.26$ and $0.36$. Hence to 1-$\sigma$ confidence
level, about three quarters of our two-dimensional
$(n,\omebar[ls]{\phi})$ parameter space can be excluded for any given
$\kappa$. For instance, without quintessence, the value of $\kappa$
lies between $0.26$ and $0.29$ for $n \in [0.8,1.2]$. The measurement
by {\it MAP} or {\it PLANCK} of a value of $\kappa>0.29$ would therefore be a
strong hint of a dark energy component playing a role at last
scattering.

\section{Model dependence} \label{sec::dependence}
The fitting formulae were obtained using a standard exponential
potential \cite{Wetterich:1988fm} for the quintessence component.
Because the shifts are almost independent of post recombination
physics, we expect the results to be approximately correct for any
realization of quintessence, i.e. all potentials. One should however
be cautious with models that are qualitatively extremely different
from the exponential potential before last scattering, as for example
the Ratra-Peebles inverse power law \cite{Peebles:1988ek} with
substantial \omebar[ls]{\phi}. In these models there is a sharp
increase in \ome[]{\phi} during recombination, whereas the
quintessence content for the exponential potential is fairly constant
at this epoch.

The inverse power law is characterized by its potential 
$V^{\rm IPL} = A / \varphi^{\alpha}$. Models with $\alpha \gtrapprox 2$ 
are phenomenologically disfavoured \cite{Balbi:2001kj}. 
We use these models only as cross checks for the fitting formulae.

In terms of phase shifts, one finds that the sensitive relative shifts
of the first trough and the second peak differ substantially for the
two models (see Table \ref{tab::leaping}).  However, $\bar\varphi$ and
$\kappa$ are seen to be more robust and the deduced value of $l_A$ is
accurate to within one percent in every case.

\begin{table*}
\begin{center}
\begin{tabular}{cccccccccc}
\toprule
\omebar[ls]{\phi} ($\%$)  & $l_1$ & $l_{3/2}$ & $l_2$ &
$l_3$ &  $l_A$ & $\bar\varphi$ &  \sh[3/2] &  \sh[2]  & $\kappa$ \\
\midrule
\multicolumn{10}{c}{Leaping kinetic term}\\[0.4ex]  

3 & 214 & 396 & 521 & 788 & 293 & 0.269 & -0.121 &  -0.045 &
0.287 \\ 
& & & & &  294 & 0.271 & -0.119 &  -0.041 & 0.292 \\[0.4ex]
13 & 210 & 396 & 522 & 799 & 301 & 0.301 & -0.120 & -0.038 &
0.317 \\ 
& & & & &  301 & 0.301 & -0.120 & -0.038 & 0.318 \\[0.4ex]
22 & 208 & 397 & 524 & 808 & 307 & 0.324 &  -0.116 & -0.030 &
0.341 \\
& &  & & &  305 & 0.320 & -0.120 & -0.035 & 0.333 \\[0.4ex]
\multicolumn{10}{c}{Ratra Peebles inverse power law}  \\[0.4ex]
$5\times 10^{-3}$ & 199 & 366  & 480 & 724 & 269 & 0.259 & -0.119 & -0.043  &
0.278 \\
& & & & &  270 & 0.261 & -0.117 & -0.038 & 0.284 \\[0.4ex] 
10 & 178 & 339 & 443 & 674 & 251 & 0.294 & -0.140 & -0.054 &
0.304 \\ 
& & & & &  253 & 0.298 & -0.138 & -0.050 & 0.312 \\[0.4ex] 
22 & 172 & 338 & 444 & 683 & 258 & 0.333 & -0.144& -0.057 &
0.340 \\ 
& & & & &  258 & 0.334 &  -0.145 & -0.057 & 0.340 \\
\bottomrule
\end{tabular}
\caption{The peak locations and the phase shifts of leaping kinetic term
  {\protect\cite{Hebecker:2001zb}} and Ratra Peebles inverse power law
  {\protect\cite{Peebles:1988ek}} 
  models for $\obh = 0.021,\ \ome[0]{\phi}=0.6,\ 
  h=0.65,\ n=1$ and varying \omebar[ls]{\phi}.
  The inverse power law models correspond to $\alpha=6,\, 22$ and $40$
  respectivly.
  The first row of each
  model gives the {\textsc CMBFAST}-obtained values of the locations of
  the peaks and the phase shifts as well as $l_A$ and $\kappa$. The
 second row gives the values deduced using the method described in 
 Section \ref{sec::phase}.} 
\label{tab::leaping}
\end{center}
\end{table*}

\section{Conclusions}
In this paper we have shown that within a wide range of parameters,
one can accurately deduce the acoustic scale $l_A$, as well as the
shifts of the peaks and troughs provided the locations of the first
three peaks are measured. Not only will this enable faster testing in
likelihood analysis by providing a filter before any fluctuation
equations are solved, but it could also in principle lead to a
detection of quintessence -- measuring a non-zero value of the dark
energy at last scattering (e.g by computing the quantity $\kappa$
described in Section \ref{sec::kappa}) would distinguish it from a
cosmological constant, whose contribution to the energy density of the
Universe would become significant only very recently.

\section*{Acknowledgments}

We would like to thank C.~Wetterich and G.~Aarts for helpful
discussions.

\appendix
\section{Fitting formulae}\label{sec::fit}

We present here our fitting formulae for the overall phase shift
$\bar{\varphi}$, followed by the relative shifts of the first trough
(\sh[3/2]) and the second (\sh[2]) and third (\sh[3])
peaks.\footnote{A small c++ package providing functions for the shifts
  is available at http://www.thphys.uni-heidelberg.de/\~\,
  \!\!\!doran/peak.html} In each case we also give an estimate of the
accuracy of the formulae.

\subsection{Overall phase shift $\bar\varphi$}
For the overall phase shift $\bar{\varphi}$ (i.e. the phase shift of
the first peak) we find the formula
\begin{equation} \label{phi_1}
\bar{\varphi} =(1.466 - 0.466n) \left[ a_1 r_{\star}^{a_2} + 0.291
  \omebar[ls]{\phi} \right],
\end{equation}
where $a_1$ and $a_2$ are given by
\begin{eqnarray}
a_1 & = & 0.286 + 0.626\left(\Omega_b h^2\right) \\
a_2 & = & 0.1786 - 6.308 \,\obh + 174.9\left(\obh\right)^2\\
&&  - 1168 \left(\obh\right)^3.
\end{eqnarray}
It contains the main dependence of any shift $\varphi_m$ on
\omebar[ls]{\phi}.  The 1-$\sigma$ error for $\bar\varphi$ is
\begin{equation}
\Delta \bar\varphi = 0.0031
\end{equation}
\subsection{Relative shift of first trough \sh[3/2]}
The relative shift of the first trough is a very sensitive quantity
spanning a wide range of values. It can very well be used to restrict
the allowed parameter space for cosmological models.  We have
\begin{equation}
\sh[3/2] = b_0 + b_1 \rstar^{1/3}  \exp(b_2 \rstar) + 0.158\, (n-1),
\end{equation}
with
\begin{eqnarray}
\nonumber b_0 &=& -0.086 - 0.079 \, \omebar[ls]{\phi} - \left
  (2.22 - 18.1 \omebar[ls]{\phi}\right ) \obh \\
&& -\left [140 + 403 \omebar[ls]{\phi} \right]
\left(\obh\right)^2\\
b_1 &=& 0.39 - 0.98 \, \omebar[ls]{\phi} - \left (18.1 - 29.2
\omebar[ls]{\phi}\right)   \obh \\
&&+ 440   \left(\obh\right)^2 \\ 
b_2 &=& -0.57-3.8 \exp\left\{-2365.0 \left(\obh\right)^2\right\}.
\end{eqnarray}
For the one standard-deviation error we have
\begin{equation}
\Delta \sh[3/2] = 0.0039.
\end{equation}

\subsection{Relative shift of second peak \sh[2]}
The relative shift of the second peak is a very sensitive quantity.
It is thus not surprising to find a strong dependence of \sh[2] on the
parameters. We have

\begin{equation}
\delta \varphi_2 = c_0 - c_1 \rstar - c_2 \rstar ^{-c_3} + 0.05\,
(n-1),
\end{equation}
with
\begin{eqnarray}
c_0 &=& -0.1 + \left( 0.213 - 0.123 \,\omebar[ls]{\phi} \right)\\
&&\times \exp\left\{ - \left( 52 - 63.6 
\,\omebar[ls]{\phi}\right) \obh \right\}\\
c_1 &=& 0.063 \,\exp\left \{-3500 \left(\obh\right)^2\right\} +
0.015\\
c_2 &=& 6\times 10^{-6} + 0.137 \left (\obh - 0.07 \right) ^2\\ 
c_3 &=& 0.8 + 2.3 \,\omebar[ls]{\phi} + \left( 70 - 126
  \,\omebar[ls]{\phi}\right) \obh.
\end{eqnarray}
The error of this approximation is
\begin{equation}
\Delta \sh[2] = 0.0044.
\end{equation}

\subsection{Relative shift of third peak $\delta \varphi_3$}
For the third peak, we find
\begin{equation}
\delta \varphi_3 = 10 - d_1\rstar^{d_2} + 0.08\, (n-1),
\end{equation}
with 
\begin{eqnarray}
d_1 &=& 9.97 + \left(3.3 -3 \ome[ls]{\phi}\right) \obh \\
\nonumber d_2 &=& 0.0016 - 0.0067 \,\ome[ls]{\phi}  + \left(0.196 - 0.22
  \,\ome[ls]{\phi}\right) \obh\\ 
&& + \frac{(2.25 + 2.77  \,\ome[ls]{\phi} ) \times 10^{-5}}{ \obh},
\end{eqnarray}
and error given by
\begin{equation}
\Delta \sh[3] = 0.0052.
\end{equation}

\subsection{Overall shift of third peak $\varphi_3$}
For completeness, we give a fit for $\varphi_3$ which in principle
could be obtained by adding $\bar\varphi$ and \sh[3]. However, a
one-step-fit yields better errors here.  Our formula is
\begin{equation}
\varphi_3 = e_1  \left(1 + e_3 \rstar\right)\rstar^{e_2} + e_4 - 0.037
\, (n-1),
\end{equation}
with
\begin{eqnarray} 
  e_1 &=& 0.302 - 2.112 \,\obh + 0.15 \exp\left\{-384 \obh\right\}\\
  e_2 &=& -0.04 - 4.5 \,\obh\\
  e_3 &=& \left(-0.118 + 44.7 \,\obh\right) \omebar[ls]{\phi} \\
  e_4 &=& \left(0.214 \exp\left\{-48 \obh\right\}+ 0.106\right)
  \omebar[ls]{\phi},
\end{eqnarray}
and error 
\begin{equation}
\Delta \varphi_3 = 0.0017. 
\end{equation}


\begin{thebibliography}{}
%\cite{Huey:1999se}
\bibitem{Huey:1999se} 
Huey, G., Wang, L.,
 Dave, R., Caldwell, R. R., Steinhardt, P. J., 1999,
%``Resolving the Cosmological Missing Energy Problem,''
Phys.\ Rev.\ D, 59, 063005 [astro-ph/9804285]
%%CITATION = ASTRO-PH 9804285;%%

%\cite{Hu:1996qz}
\bibitem{Hu:1996qz}
Hu, W., White, M., 1996,
%``Measuring the Curvature of the Universe,''
in Proceedings of 31st Rencontres de Moriond: Microwave
  Background Anisotropies, Les Arcs, France, 16-23 March 1996
(Editions Frontieres) [astro-ph/9606140]
%%CITATION = ASTRO-PH 9606140;%%

%\cite{Amendola:2000er}
\bibitem{Amendola:2000er}
Amendola, L., 2000,
%``Coupled quintessence,''
Phys.\ Rev.\ D, 62, 043511 [astro-ph/9908023]
%%CITATION = ASTRO-PH 9908023;%%

%\cite{Brax:2000yb}
\bibitem{Brax:2000yb} 
Brax, P., Martin, J., Riazuelo, A., 2000,
%``Exhaustive study of cosmic microwave background anisotropies in
%quintessential scenarios,''
Phys.\ Rev.\ D, 62, 103505 [astro-ph/0005428]
%%CITATION = ASTRO-PH 0005428;%%

%\cite{Coble:1997te}
\bibitem{Coble:1997te} 
Coble, K., Dodelson, S., Frieman, J. A., 1997,
%``Dynamical Lambda models of structure formation,''
Phys.\ Rev.\ D, 55, 1851 [astro-ph/9608122].
%%CITATION = ASTRO-PH 9608122;%%

%\cite{Doran:2000jt}
\bibitem{Doran:2000jt} 
Doran, M., Lilley, M. J., Schwindt. J., Wetterich, C., 2000,
%``Quintessence and the separation of CMB peaks,''
ApJ in press [astro-ph/0012139].
%%CITATION = ASTRO-PH 0012139;%%


%\cite{Seljak:1996is}
%\bibitem[\protect\citename{Seljak \& Zaldarriaga
%}{1996}]{Seljak:1996is} 
\bibitem{Seljak:1996is} 
Seljak, U., Zaldarriaga, M., 1996,
%``A Line of Sight Approach to Cosmic Microwave Background Anisotropies,''
ApJ, 469, 437 [astro-ph/9603033]
%%CITATION = ASTRO-PH 9603033;%%




%\cite{Kamionkowski:1994aw}
\bibitem{Kamionkowski:1994aw}
Kamionkowski, M., Spergel, D. N., Sugiyama, N., 1994,
%``Small scale cosmic microwave background anisotropies as a probe of
%the geometry of the universe,''
ApJ, 426, L57 [astro-ph/9401003]
%%CITATION = ASTRO-PH 9401003;%%

%\cite{Weinberg:2000ts}
\bibitem{Weinberg:2000ts}
Weinberg, S., 2000,
%``Curvature dependence of peaks in the cosmic microwave background
%distribution,'' 
Phys.\ Rev.\ D, 62, 127302 [astro-ph/0006276]
%%CITATION = PHRVA,D62,127302;%%


  
%\cite{Hu:1997qs}
\bibitem{Hu:1997qs}
Hu, W., Sugiyama, N., Silk, J., 1997,
%``The Physics of microwave background anisotropies,''
Nat, 386, 37 [astro-ph/9604166]
%%CITATION = ASTRO-PH 9604166;%%



%\cite{Hu:2000ti}
\bibitem{Hu:2000ti}
Hu, W., Fukugita, M., Zaldarriaga, M., Tegmark, M., 2000,
%``CMB Observables and Their Cosmological Implications,''
astro-ph/0006436
%%CITATION = ASTRO-PH 0006436;%%


%\cite{Hu:1995uz}
\bibitem{Hu:1995uz}
Hu, W., Sugiyama, N., 1995,
%``Anisotropies in the cosmic microwave background: An Analytic approach,''
ApJ, 444, 489
%%CITATION = ASJOA,444,489;%%


%\cite{Wetterich:1988fm}
\bibitem{Wetterich:1988fm}
Wetterich, C., 1988,
%``Cosmology And The Fate Of Dilatation Symmetry,''
Nucl.\ Phys.\ B, 302, 668 
%%CITATION = NUPHA,B302,668;%%



%\cite{Peebles:1988ek}
\bibitem{Peebles:1988ek} 
Peebles, P. J. E., Ratra, B., 1988,
%``Cosmology With A Time Variable Cosmological 'Constant',''
ApJ, 325, L17
%%CITATION = ASJOA,325,L17;%%




%\cite{Balbi:2001kj}
\bibitem{Balbi:2001kj}
A.~Balbi, C.~Baccigalupi, S.~Matarrese, F.~Perrotta and N.~Vittorio,
%``Implications on quintessence models from MAXIMA-1 and BOOMERANG-98,''
Astrophys.\ J.\  {\bf 547} (2001) L89
%[astro-ph/0009432].
%%CITATION = ASTRO-PH 0009432;%%

%\cite{Hebecker:2001zb}
\bibitem{Hebecker:2001zb} 
Hebecker, A., Wetterich, C., 2001,
%``Natural quintessence?,''
Phys.\ Lett.\ B, 497, 281
[hep-ph/0008205]
%%CITATION = HEP-PH 0008205;%%


\end{thebibliography}
\end{document}